\documentclass[prl,twocolumn,amsmath,superscriptaddress,amssymb]{revtex4}

\usepackage{graphicx}
\usepackage{dcolumn}
\usepackage{bm}
\usepackage{amssymb}
\usepackage{amsmath}
\usepackage{wasysym}
\usepackage{color}
\usepackage{times}

\usepackage{placeins}
\usepackage{epstopdf}

\newcommand{\ket}[1]{\left|\,#1\,\right\rangle}

\usepackage[normalem]{ulem}

\makeatletter
\renewcommand*\env@matrix[1][c]{\hskip -\arraycolsep
  \let\@ifnextchar\new@ifnextchar
  \array{*\c@MaxMatrixCols #1}}
\makeatother
\begin{document}

\title{Orbital-driven Rashba effect in a binary honeycomb monolayer AgTe}

\author{Maximilian \"Unzelmann}\affiliation{Experimentelle Physik VII and W\"urzburg-Dresden Cluster of Excellence ct.qmat, Universit\"at W\"urzburg, Am Hubland, D-97074 W\"urzburg, Germany}
\author{Hendrik Bentmann}\email{Hendrik.Bentmann@physik.uni-wuerzburg.de}\affiliation{Experimentelle Physik VII and W\"urzburg-Dresden Cluster of Excellence ct.qmat, Universit\"at W\"urzburg, Am Hubland, D-97074 W\"urzburg, Germany}
\author{Philipp Eck}\affiliation{Theoretische Physik I and and W\"urzburg-Dresden Cluster of Excellence ct.qmat, Universit\"at W\"urzburg, D-97074 W\"urzburg, Germany}
\author{Tilman Ki{\ss}linger}\affiliation{Lehrstuhl f\"ur Festk\"orperphysik, Universit\"at Erlangen-N\"urnberg, Staudtstra{\ss}e 7, D-91058 Erlangen, Germany}
\author{Begmuhammet Geldiyev}\affiliation{Lehrstuhl f\"ur Festk\"orperphysik, Universit\"at Erlangen-N\"urnberg, Staudtstra{\ss}e 7, D-91058 Erlangen, Germany}
\author{Janek Rieger}\affiliation{Lehrstuhl f\"ur Festk\"orperphysik, Universit\"at Erlangen-N\"urnberg, Staudtstra{\ss}e 7, D-91058 Erlangen, Germany}
\author{Simon Moser}\affiliation{Experimentelle Physik IV and W\"urzburg-Dresden Cluster of Excellence ct.qmat, Universit\"at W\"urzburg, Am Hubland, D-97074 W\"urzburg, Germany}
\author{Raphael C. Vidal}\affiliation{Experimentelle Physik VII and W\"urzburg-Dresden Cluster of Excellence ct.qmat, Universit\"at W\"urzburg, Am Hubland, D-97074 W\"urzburg, Germany}
\author{Katharina Ki{\ss}ner}\affiliation{Experimentelle Physik VII and W\"urzburg-Dresden Cluster of Excellence ct.qmat, Universit\"at W\"urzburg, Am Hubland, D-97074 W\"urzburg, Germany}
\author{Lutz Hammer}\affiliation{Lehrstuhl f\"ur Festk\"orperphysik, Universit\"at Erlangen-N\"urnberg, Staudtstra{\ss}e 7, D-91058 Erlangen, Germany}
\author{M. Alexander Schneider}\affiliation{Lehrstuhl f\"ur Festk\"orperphysik, Universit\"at Erlangen-N\"urnberg, Staudtstra{\ss}e 7, D-91058 Erlangen, Germany}
\author{Thomas Fauster}\affiliation{Lehrstuhl f\"ur Festk\"orperphysik, Universit\"at Erlangen-N\"urnberg, Staudtstra{\ss}e 7, D-91058 Erlangen, Germany}
\author{Giorgio Sangiovanni}\affiliation{Institut f\"ur Theoretische Physik und Astrophysik and W\"urzburg-Dresden Cluster of Excellence ct.qmat, Universit\"at W\"urzburg, 97074 W\"urzburg, Germany}
\author{Domenico Di Sante}\affiliation{Theoretische Physik I and and W\"urzburg-Dresden Cluster of Excellence ct.qmat, Universit\"at W\"urzburg, D-97074 W\"urzburg, Germany}
\author{Friedrich Reinert}\affiliation{Experimentelle Physik VII and W\"urzburg-Dresden Cluster of Excellence ct.qmat, Universit\"at W\"urzburg, Am Hubland, D-97074 W\"urzburg, Germany}

\date{\today}

\begin{abstract}
The Rashba effect is fundamental to the physics of two-dimensional electron systems and underlies a variety of spintronic phenomena. It has been proposed that the formation of Rashba-type spin splittings originates microscopically from the existence of orbital angular momentum (OAM) in the Bloch wave functions. Here, we present detailed experimental evidence for this OAM-based origin of the Rashba effect by angle-resolved photoemission (ARPES) and two-photon photoemission (2PPE) experiments for a monolayer AgTe on Ag(111). Using quantitative low-energy electron diffraction (LEED) analysis we determine the structural parameters and the stacking of the honeycomb overlayer with picometer precision. Based on an orbital-symmetry analysis in ARPES and supported by first-principles calculations, we unequivocally relate the presence and absence of Rashba-type spin splittings in different bands of AgTe to the existence of OAM.   
     
\end{abstract}
\maketitle

The breaking of inversion symmetry in solids allows for the formation of spin-polarized electronic states via spin-orbit interaction, even in the absence of magnetism \cite{Rashba:84,Manchon2015}. This effect is fundamental to many key phenomenona in solid state physics, such as the spin-momentum-locked surface states in topological insulators \cite{Konig2007,Hasan2010}, the realization of Weyl fermions in non-centrosymmetric crystals \cite{Armitage:18} and the intrinsic spin Hall effect \cite{Sinova:04}. At surfaces and interfaces, where inversion-symmetry breaking (ISB) is inherently present, Rashba-type spin splittings of two-dimensional (2D) electron states \cite{Rashba:84} have been observed in many material systems \cite{LaShell:96,Nitta:97,Hoesch:04,Koroteev:04,Ast2007,Dil:09,Caviglia:10,ishizaka:11,King:11,marchenko:12,Crepaldi:12, Krasovskii2015,Eickholt:16,Sakamoto:16,Niesner2016,Eickholt:18,Rosenzweig:18} and are expected to underlie a variety of spintronic phenomena \cite{Manchon2015}, such as spin-charge conversion \cite{lesne:16} and spin-orbit torques \cite{kurebayashi:14}. Understanding the microscopic nature of the interplay between ISB and spin-orbit coupling (SOC) in generating large spin splittings therefore has broad relevance in condensed matter physics and materials science \cite{DiSante:13,Manchon2015,Sunko2017}.       

Rashba-type spin splittings at surfaces are usually accurately described by first-principles calculations \cite{Nicolay:01,Henk:03,Koroteev:04,bihlmayer:06,Bihlmayer2007,nagano:09,Eremeev:12,Chiang:13,Krasosvkii:14,Ishida:14,Krasovskii2015,Nechaev:19}, which also provide immediate insight into the underlying physical mechanisms. It has been predicted theoretically that the formation of chiral orbital angular momentum (OAM) in the Bloch wave functions --as a consequence of ISB and independently of SOC-- is a generic effect that precedes the actual spin splitting due to SOC and crucially influences its magnitude \cite{Park2011,Park2012,Go2017}. This scenario has the profound implication that the main driving force for the Rashba effect is not the electron's spin but rather its orbital dipole moment \cite{Park2011,Park2012,Go2017}. Experimentally, signatures of the existence of OAM for different spin-orbit split surface states have been observed by circular dichroism in angle-resolved photoelectron spectroscopy (ARPES) \cite{Park2012,Taniguchi:12,Kim:12,Kim:13}. However, systematic experimental evidence for an OAM-based origin of the Rashba effect, namely that the presence or absence of OAM is directly related to the existence of Rashba splitting, is still lacking.          

Here, we investigate the microscopic origin of the Rashba effect by means of ARPES experiments and density functional theory (DFT) calculations for the 2D honeycomb monolayer AgTe grown on Ag(111) \cite{Liu2019}. The valence band of the AgTe monolayer is formed by two bands that derive mainly from Te $5p_{x,y}$ in-plane orbitals. In spite of their same atomic-orbital origin and their same spatial localization directly at the surface, these two bands show vastly different Rashba-type spin splittings, namely vanishing for one band and comparable to the giant splittings observed in Bi-compounds \cite{Ast2007,Bentmann2009,Rosenzweig:18} for the other. Our measurements and calculations show how this surprising behavior is related to different in-plane wave-function symmetries of the two bands that either allow or inhibit the formation of OAM. Only the OAM-carrying band is found to develop a sizable Rashba-type spin splitting, which provides conclusive experimental evidence for the predicted OAM-based microscopic origin of the Rashba effect \cite{Park2011,Park2012}.

\begin{figure}[t]
\includegraphics[width=\columnwidth]{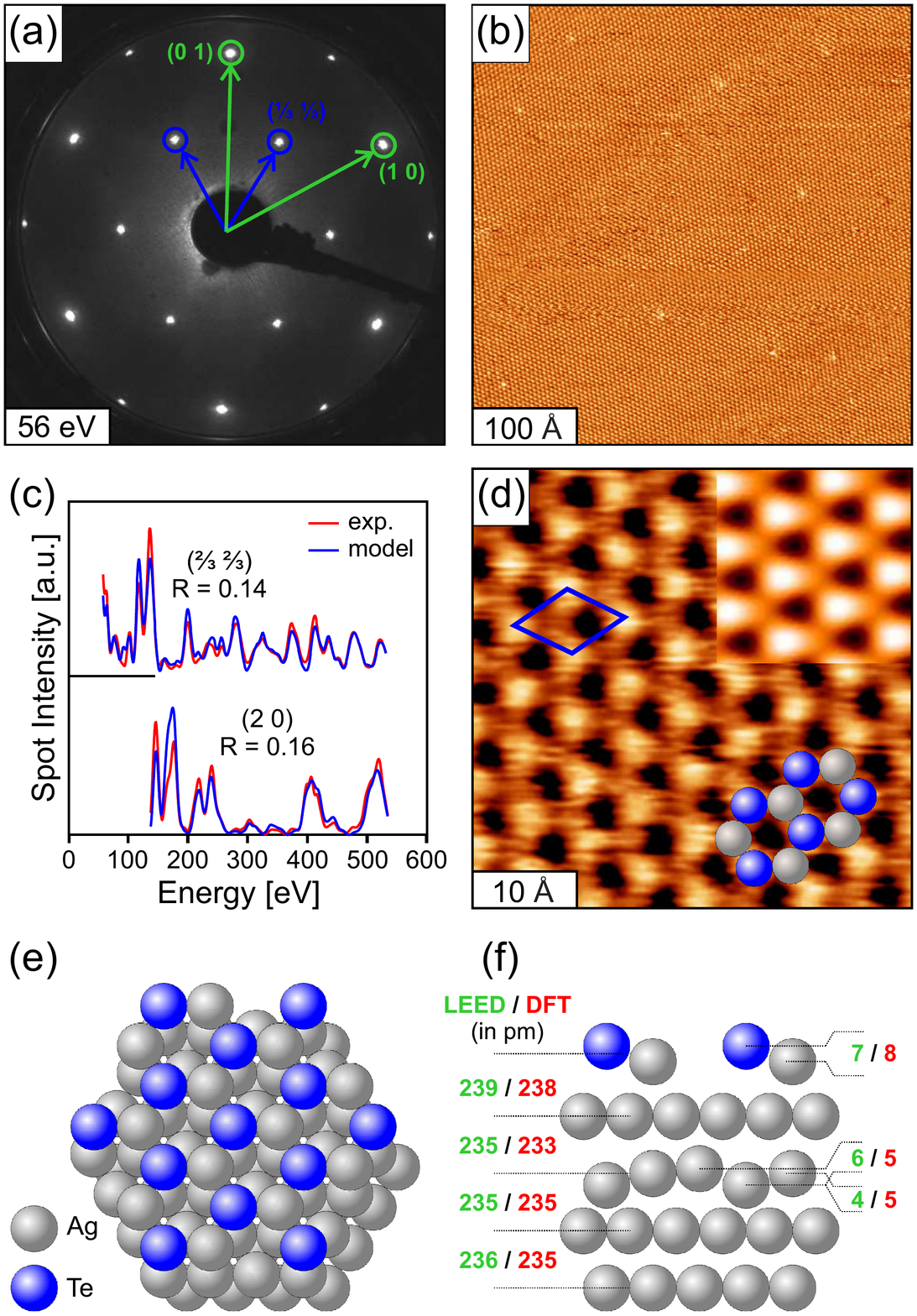}
\caption{(a) LEED image of the AgTe honeycomb lattice on Ag(111) forming a $\left( \sqrt{3} \times \sqrt{3}\right)-R30^{\circ}$-Te superstructure.
(b) Corresponding large-scale STM image (tunneling parameters: $U_b =$~80~mV, $I_t =$~1.0~nA) showing a defect density below 1~$\%$.
(c) Comparison of selected LEED-I-V spectra (red) and best-fit curves (blue). The single beam R-factors of the shown spectra are close to the overall fit quality of R = 0.15.
(d) Atomically resolved STM image ($U_b =$~100~mV, $I_t =$~1.0~nA) of the AgTe honeycomb lattice together with the simulated STM image as obtained by our DFT calculations. The bright protrusions are clearly the slightly outward buckled Te atoms.
(e) Top view of the surface structure showing the hcp-like stacking of the outermost surface layer as determined from LEED-I-V.
(f) Side view with geometrical parameters as obtained from the LEED-I-V analysis (green) and DFT (red), respectively.}
\label{fig1}
\end{figure}

We performed ARPES, two-photon-photoemission (2PPE) \cite{Rosenzweig:18}, low-energy-electron-diffraction (LEED) and scanning-tunneling-microscopy (STM) experiments as well as calculations based on density functional theory (DFT), as described in the Supplemental Material (SM) \cite{Supp1}. The AgTe layer was prepared by Te deposition on a Ag(111) substrate (see SM for a detailed description \cite{Supp1}). This procedure led to a LEED-pattern as depicted in Fig. 1(a) showing a well-ordered $(\sqrt 3 \times \sqrt 3)R30^\circ$ superstructure. 
The high degree of order is corroborated by STM measurements like the one in Fig.~1(b) that show a
 fully covered surface with a defect density below 1~\%.

The atomically resolved STM image of Fig.~1(d) shows a honeycomb arrangement of the surface atoms, in agreement with recent findings \cite{Liu2019}. However, our quantitative LEED-IV analysis proves at a Pendry R-factor 0.15 that this is indeed a AgTe honeycomb arrangement of the first surface layer in hcp-hollow sites but contrary to Ref.~\cite{Liu2019} it is not flat but shows a noticable buckling.
All other tested alternative models (fcc-honeycomb, Te adlayer, substitutional Ag$_2$Te surface alloy)
 could be safely excluded as they resulted in R-factors of R $>$ 0.4. 
The excellent agreement between the experiment and the model calculations is visualized in Fig.~1(c) by two exemplary beams. 
The crystallographic parameters determined by the LEED analysis
 are depicted in the ball model in Fig. 1(f). Our DFT calculations match within 1.5 pm for all values including the buckling of the AgTe layer. Our DFT-STM simulations [inset in Fig. 1(d)] show that Te is imaged brighter than Ag in STM in contrast to the propositions of~\cite{Liu2019}.

\begin{figure*}[t]
\includegraphics[width=\textwidth]{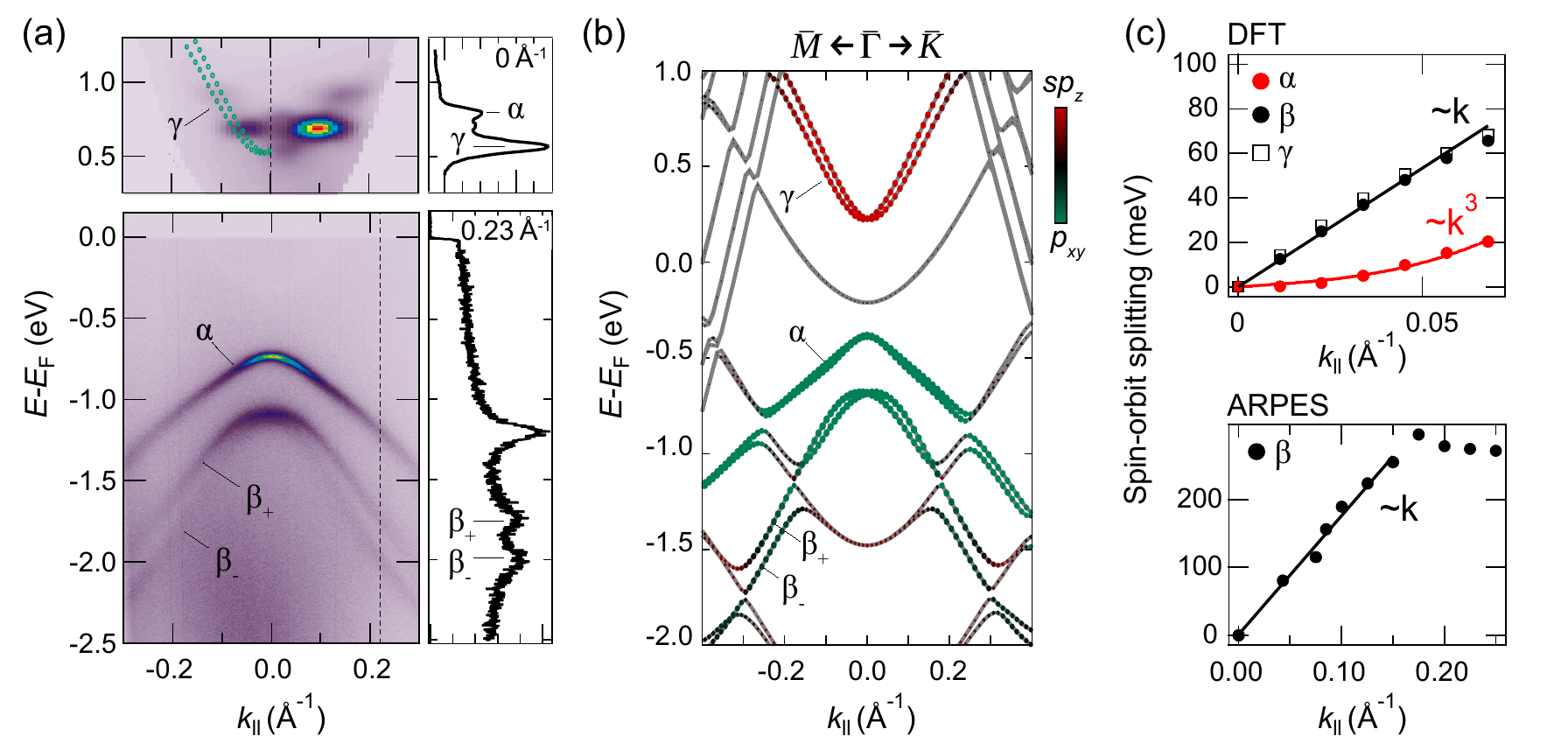}
\caption{Band structure of an AgTe monolayer on Ag(111). (a) Angle-resolved photoemission ($h\nu =$~21.21$\,$eV) and two-photon photoemission ($h\nu =$~1.55 + 4.65$\,$eV) data sets measured along $\bar{\Gamma}\bar{K}$ and $\bar{\Gamma}\bar{M}$, respectively. Two hole-like bands, $\alpha$ and $\beta$, are observed in the occupied regime below $E_F$ and one electron-like band, $\gamma$, is found above $E_F$. The band $\beta$ shows a sizable Rashba-type spin splitting into the branches $\beta_\pm$, as also visible in the energy distribution curve (EDC) at a $k_\parallel$ marked by the dashed line. (b) DFT band structure calculation for AgTe/Ag(111). The size of the dots indicates the localization of the states in the AgTe monolayer. (c) Top panel: spin-orbit splitting of $\alpha$, $\beta$ and $\gamma$ in dependence of $k_\parallel$ as extracted from the DFT calculation in (b). Bottom panel: Measured spin-orbit splitting of $\beta$ as determined from the data in (a). Lines in the top and bottom panels are linear or cubic fits to the data points as labeled.}
\label{fig2}
\end{figure*}

Based on the quantitative structure determination of the AgTe honeycomb layer we will discuss its electronic properties in the following. Our ARPES and 2PPE data reveal two hole-like bands $\alpha$ and $\beta$ below the Fermi level $E_F$ and another electron-like band $\gamma$ above $E_F$ [Fig.~\ref{fig2}(a)], in excellent agreement with our calculations in Fig.~\ref{fig2}(b). In the 2PPE energy distribution curve (EDC) at $k_\parallel =0$ [top right of Fig. 2(a)] $\gamma$ is seen as intermediate state and $\alpha$ as initial state. For $k_\parallel \approx0.1\mathrm{\AA}^{-1}$ resonant transitions between the two bands are observed. A closer inspection reveals a sizable Rashba-type spin splitting for $\beta$ into two branches $\beta_{+}$ and $\beta_{-}$, which is also seen in the EDC in Fig.~\ref{fig2}(a). In accordance with the Rashba-model the splitting is linear in $k_\parallel$ near the $\bar{\Gamma}$-point with a Rashba parameter $\Delta_{\beta}=0.88\pm 0.02\,\mathrm{eV}\mathrm{\AA}$ [Fig.~\ref{fig2}(c)]. For $k_\parallel \geq 0.16\,\mathrm{\AA}$ the splitting remains nearly constant indicating deviations from the Rashba model at higher wave vectors. 

The Rashba parameter for the band $\beta$ in AgTe is thus remarkably large. It is of similar magnitude as for the structurally related systems Ag$_2$Pb \cite{Grioni:06,el-kareh:14} and Cu$_2$Bi \cite{Bentmann2009}, despite their significantly larger atomic SOC, and much larger than for Ag$_2$Sb \cite{Moreschini:09} and Ag$_2$Sn \cite{Uhrberg:13}, with comparable atomic SOC strength. In stark contrast, for $\alpha$ the data do not show any observable spin splitting. Our calculations confirm a strong difference in the spin splitting for $\alpha$ and $\beta$, as already apparent qualitatively in Fig.~\ref{fig2}(b). We restrict the quantitative analysis of the calculated splitting in Fig.~\ref{fig2}(c) to a narrow region around the $\bar{\Gamma}$-point. At higher $k_\parallel$ the spin splitting gets distorted by hybridization with quantum-well states, originating from the finite size of the employed slab \cite{Supp1}. As can be seen in Fig.~\ref{fig2}(c) the calculated spin splitting for $\beta$ is indeed much larger than for $\alpha$. It is also linear in $k_\parallel$ whereas the one for $\alpha$ shows a $k_\parallel^{3}$ dependence. This indicates that the small spin splitting of $\alpha$ arises from higher-order contributions \cite{Vajna:12,Nechaev:19}, and that the linear-in-$k_\parallel$ Rashba-term is negligibly small, i.e. $\Delta_{\alpha}\approx0$. 

\begin{figure}[t]
\includegraphics[width=\columnwidth]{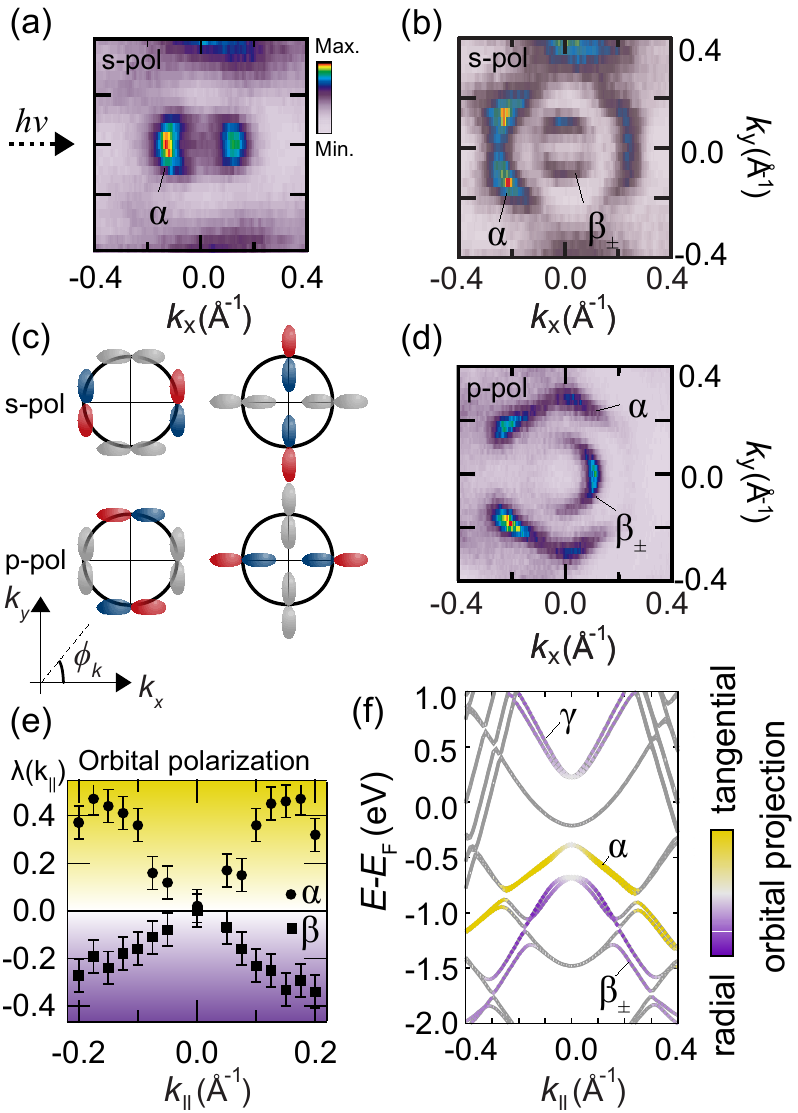}
\caption{Orbital composition of the band structure. Constant energy cuts (CEC) of the measured band structure taken at $E-E_F = -950\,$meV (a) and $-1300\,$meV (b) with s-polarized light at $h\nu=65\,$eV. (c) Sketch of tangential $p_t$ and radial $p_r$ in-plane orbital configurations. Colored (grey) orbitals indicate an enhanced (suppressed) ARPES intensity in a CEC for s- and p-polarized light, as expected from dipole-selection rules (see text for details). (d) CEC at $-1300\,$meV obtained with p-polarized light at $h\nu=25\,$eV. The plane of light-incidence for all measurements is the $xz$-plane as indicated by the arrow in (a). (e) Intensity asymmetry parameter $\lambda(k_\parallel)$ as extracted from ARPES data measured with with s-polarized light (see text for details). (f) DFT calculation of the AgTe band structure projected onto tangential and radial Te $p$ orbitals.}
\label{fig3}
\end{figure}

To explain the strong difference in $\Delta_{\alpha}$ and $\Delta_{\beta}$ we first consider a simple model, in close analogy to the tight-binding approaches in Refs.~\cite{Petersen2000,Park2012,Go2017,Hong:19}. Our DFT calculations show that $\gamma$ is mainly of $sp_z$ orbital character while $\alpha$ and $\beta$ are of $p_{xy}$ character [Fig.~\ref{fig2}(b)]. The mixing between Ag $s$ and Te $p_z$ orbitals is induced by ISB with a characteristic energy scale $\Delta_{ISB}$ \cite{Go2017}. The two $p_{xy}$ bands can be expressed as one radially aligned state $p_{r}$ and one tangentially aligned state $p_{t}$ \cite{Park2012}, see also Figs.~\ref{fig3}(c) and (f). Note that in the case of pure $sp_z$, $p_{r}$ and $p_{t}$ orbital characters all three bands would have vanishing OAM. A decisive role for the formation of OAM is played by the hopping term $\delta_{sp}$ which couples $s$ and $p_{r}$ orbitals, yielding eigenstates of the form $\ket{p_r} + ik_\parallel \delta_{sp}\ket{sp_z}$ for $\beta$ and $\ket{sp_z} + ik_\parallel \delta_{sp}\ket{p_r}$ for $\gamma$ \cite{Go2017}, see Sect. IV of \cite{Supp1} for more details. The orbital mixing induces a finite OAM $L \propto k_\parallel \delta_{sp} \Delta_{ISB}$ for these two states \cite{Go2017,Supp1}. Importantly, $\ket{p_{t}}$ is not influenced by ISB and $\delta_{sp}$ because its odd symmetry inhibits hybridization with the even $sp_z$ and $p_{r}$ orbitals in the absence of SOC, implying $L = 0$ for $\ket{p_{t}}$. The effect of SOC is considered on the basis of an atomic-like term $H_{SOC}= \zeta \mbox{\textbf{L}}\cdot \mbox{\textbf{s}} $. In the case of the two mixed states $H_{SOC}$ induces a Rashba-type spin splitting $\Delta \propto k_\parallel \zeta L$ while for the $p_{t}$ band one finds $\Delta = 0$ due to the vanishing OAM \cite{Park2012}. 

The above analysis shows that among $\alpha$, $\beta$ and $\gamma$ one can expect one band with vanishing OAM and vanishing Rashba splitting as well as two bands with a linear-in-$k_\parallel$ Rashba splitting that scales with the $\delta_{sp}$-induced OAM and the atomic SOC strength $\zeta$. The two cases are distinguished by the in-plane orbital symmetry of the wave functions, such that bands with $p_{r}$ character show Rashba splitting and bands with $p_{t}$ character do not. Our DFT calculations indeed reveal a situation closely resembling the above model considerations [see also Figs.~S2-S4 in the SM]. Figure~\ref{fig3}(f) shows the band structure of AgTe projected on the in-plane orbital contributions. We find that the band $\alpha$ is of $p_t$ character and, as discussed above, shows a vanishing linear Rashba splitting while $\beta$ and $\gamma$ carry $p_r$ character and show large linear Rashba splittings of almost the same magnitude [cf. Fig~\ref{fig2}(c)].  

To confirm experimentally the calculated orbital symmetries of the bands we performed ARPES measurements with linearly polarized light [Fig.~\ref{fig3}]. The two different in-plane orbital configurations $p_t$ and $p_{r}$ can be efficiently distinguished by evaluating intensity momentum distributions measured in s-polarized geometry \cite{Cao2013,Bawden2015}. Assuming an isotropic dispersion, based on the dipole selection rules one expects a $\cos^2{\phi_{k}}$ dependence of the intensity for a tangential-orbital state, where $\phi_{k}$ is the azimuthal angle along a constant energy contour and $\phi_{k} = 0$ corresponds to the plane of light incidence [Fig.~\ref{fig3}(c)]. In contrast, for a radial-orbital state one expects a $\sin^2{\phi_{k}}$ dependence. Our measurements for the band $\alpha$ show a maximal intensity for $\phi_{k} = 0$ and a nearly fully suppressed intensity for $\phi_{k} = \frac{\pi}{2}$ [Fig.~\ref{fig3}(a)], confirming the $p_t$ character of $\alpha$ predicted by our calculations. For $\beta$ we observe the opposite behavior with a maximum intensity at $\phi_{k} = \frac{\pi}{2}$ [Fig.~\ref{fig3}(b)], confirming the predicted $p_r$ character. Note that in the ARPES data set underlying Fig.~\ref{fig3} the Rashba splitting of the band $\beta$ is less well-resolved than in Fig.~\ref{fig2} [see also Fig.~S1 in the SM for a direct comparison], which is attributed to a reduced sample quality and energy resolution. A separate analysis of the spin-split branches is, however, not essential in the present case as our calculations predict the same orbital character. 

For a more systematic analysis we consider the asymmetry parameter $\lambda(k_\parallel)=(I_0-I_{\frac{\pi}{2}})/(I_{0}+I_{\frac{\pi}{2}})$, where $I_0$ and $I_{\frac{\pi}{2}}$ are the intensities for a given band at the wave vector $k_\parallel$ and $\phi_{k}= 0$ and $\frac{\pi}{2}$, respectively [Fig.~\ref{fig3}(e)]. $\lambda(k_\parallel)$ can be regarded as a qualitative measure of the in-plane orbital polarization, i.e. comparable to the calculation in Fig.~\ref{fig3}(f), and it was used in previous works to distinguish between $p_t$ and $p_{r}$ orbital configurations \cite{Cao2013,Bawden2015}. For s-polarized light one expects a positive $\lambda(k_\parallel)$ for $p_t$-dominated bands and a negative $\lambda(k_\parallel)$ for $p_{r}$-dominated bands. In agreement with the calculation in Fig.~\ref{fig3}(f), we observe $p_t$ character for $\alpha$ and $p_{r}$ character for $\beta$ over the entire accessible $k_\parallel$ range. Note also that the measured $\lambda(k_\parallel)$ drops towards the $\bar{\Gamma}$-point which is similarly seen in the calculation.

\begin{figure}[t]
\includegraphics[width=\columnwidth]{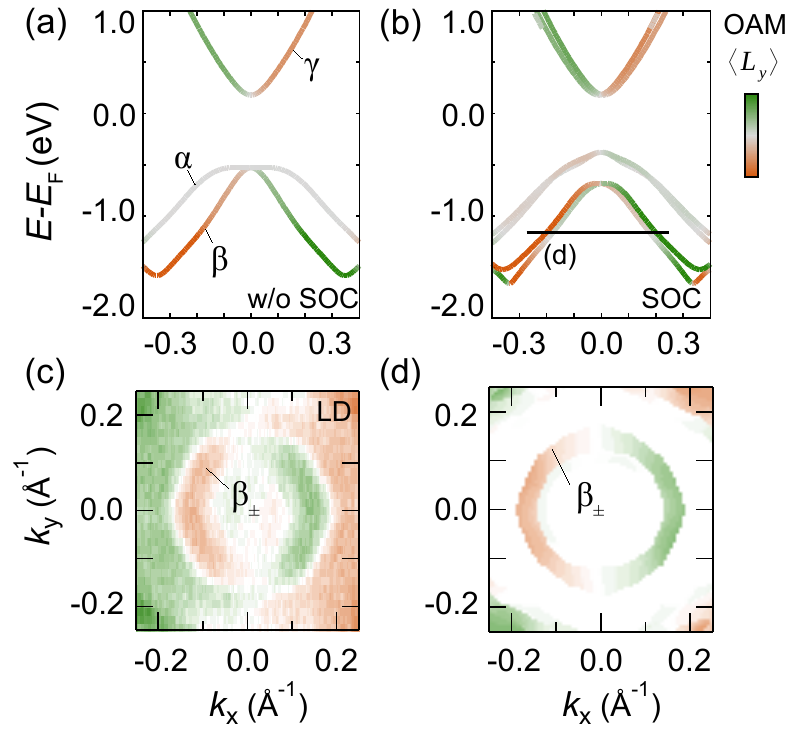}
\caption{Orbital angular momentum (OAM) in the band structure of AgTe \cite{Supp1}. (a)-(b) Calculated $\left\langle L_y\right\rangle$ OAM-projected bands in AgTe as obtained with and without including spin-orbit coupling (SOC). The green/orange color code refers to positive/negative $\left\langle L_y\right\rangle$ OAM. (c) Linear dichroism $I(k_x,k_y)-I(-k_x,k_y)$ obtained from the constant energy cut (CEC) in Fig.~\ref{fig3}(d). (d) Calculated $\left\langle L_y\right\rangle$ OAM on a CEC at an energy indicated in (b).
}
\label{fig4}
\end{figure} 

Our calculations in Figs.~\ref{fig4}(a)-(b) confirm the direct relation between the in-plane orbital symmetry and the existence of OAM, as expected from our model considerations. The bands $\beta$ and $\gamma$ show sizable OAM that increases with $k_\parallel$, while the OAM of $\alpha$ vanishes. As in the model, the OAM is already present in the absence of SOC [Fig~\ref{fig4}(a)] and only small changes are seen upon including SOC in the calculation [Fig~\ref{fig4}(b)]. 

It has been shown that circular dichroism (CD) in ARPES can provide a spectroscopic signature of OAM in two-dimensional electron systems \cite{Park2012,Taniguchi:12,Kim:12,Sunko2017}. Following this idea we consider here measurements with p-polarized light [Fig.~\ref{fig3}(d) and~\ref{fig4}(c)]. In the plane of light incidence, i.e. the $xz$ plane with $\phi_{k}= 0$, p-polarized light is expected to couple to the even $p_r$ and $sp_z$ orbitals \cite{Cao2013}. Accordingly we find that the spectral weight of $\alpha$ is suppressed [Fig.~\ref{fig3}(c)-(d)] for $\phi_{k}= 0$ while $\beta$ develops a high intensity, supporting our conclusions based on the data obtained with s-polarized light. Moreover, we observe an intensity asymmetry at $\pm k_x$ for $\beta$ [\ref{fig4}(c)]. This pronounced linear dichroism (LD) indicates a mixed $sp_z$ and $p_r$ orbital character for $\beta$ \cite{Henk:04,Bentmann2017,Min2019}, and, thus, can be tentatively regarded as an indicator for the presence of OAM, similar to the CD \cite{Park2012,Taniguchi:12,Kim:12,Sunko2017} (see sections V. and VI. of the SM). Indeed, the LD shows a similar momentum-space structure as the calculated OAM, as seen in Figs. \ref{fig4}(c)-(d).  

To conclude, our experimental and theoretical findings for an AgTe honeycomb monolayer yield comprehensive evidence for an OAM-based origin of the Rashba effect, in agreement with previous predictions \cite{Park2011,Park2012}. The key observation is that the character of the in-plane orbital texture, which is accessible experimentally from ARPES, is directly linked to the formation of chiral OAM and Rashba-type spin splitting. The magnitude of the OAM in AgTe is determined by an interplay of ISB and the kinetic hopping term $\delta_{sp}$, which couples Ag $s$ and Te in-plane $p$ orbitals on the two sublattices of the honeycomb layer. Such a kinetic energy-driven formation of OAM is strikingly similar to the mechanism underlying the recently discovered large spin splittings at oxide surfaces \cite{Sunko2017}. Based on our results for AgTe, this mechanism --the cooperation of ISB and large interatomic hopping amplitudes to generate sizable spin splittings-- will likely be applicable to a wide range of 2D monolayer materials.

\section{Acknowledgments}
We would like to thank Chris Jozwiak, Aaron Bostwick and Eli Rotenberg for experimental support during the measurements at ALS.
The work was supported by the Deutsche Forschungsgemeinschaft (DFG, German Research Foundation) – Project-ID 258499086 – SFB 1170 and through the W\"urzburg-Dresden Cluster of Excellence on Complexity and Topology in Quantum Matter – ct.qmat Project-ID 39085490 - EXC 2147. The authors acknowledge the Gauss Centre for Supercomputing e.V. for funding this project by providing computing time on the GCS Supercomputer SuperMUC at Leibniz Supercomputing Centre.

\end{document}